\documentclass[aps,pre,superscriptaddress,showpacs,twocolumn]{revtex4}

\usepackage{amsmath}
\usepackage{amssymb}
\usepackage{amsfonts}
\usepackage{mathrsfs}
\usepackage{graphicx}
\usepackage[english,british]{babel}

\begin{document}

\selectlanguage{\british}

\title{Barrier crossing driven by L{\'e}vy noise: Universality and
the Role of Noise Intensity}

\author{Aleksei V. Chechkin}
\affiliation{Institute for Theoretical Physics NSC KIPT, Akademicheskaya
st.1, 61108 Kharkov, Ukraine}
\author{Oleksii Yu. Sliusarenko}
\affiliation{Institute for Theoretical Physics NSC KIPT, Akademicheskaya
st.1, 61108 Kharkov, Ukraine}
\author{Ralf Metzler}
\affiliation{Dept of Physics, University of Ottawa - 150 Louis Pasteur,
Ottawa, ON, K1N 6N5, Canada}
\affiliation{NORDITA---Nordic Institute for Theoretical Physics,
Blegdamsvej 17, 2100 Copenhagen {\O}, Denmark}
\author{Joseph Klafter}
\affiliation{School of Chemistry, Tel Aviv University, Ramat Aviv,
Tel Aviv 69978, Israel}

\begin{abstract}
We study the barrier crossing of a particle driven by white symmetric L{\'e}vy
noise of index $\alpha$ and
intensity $D$ for three different generic types of potentials:
(a) a bistable potential; (b) a metastable potential; and (c) a truncated
harmonic potential. For the low noise intensity regime we recover the
previously proposed algebraic dependence on $D$ of the characteristic
escape time, $T_{\mathrm{esc}}\simeq C(\alpha)/D^{\mu(\alpha)}$, where
$C(\alpha)$ is a coefficient. It is shown that the exponent $\mu(\alpha)$
remains approximately constant, $\mu\approx 1$ for $0<\alpha<2$; at
$\alpha=2$ the power-law form of $T_{\mathrm{esc}}$ changes into the known
exponential dependence on $1/D$; it exhibits a
divergence-like behavior as $\alpha$ approaches 2. In this regime we observe
a monotonous increase of the escape time $T_{\mathrm{esc}}$ with increasing
$\alpha$ (keeping the noise intensity $D$ constant). The probability density
of the escape time decays exponentially. In addition, for low noise intensities
the escape times correspond to barrier crossing by multiple L{\'e}vy steps.
For high noise intensities, the escape time curves collapse for all values
of $\alpha$. At intermediate noise intensities, the escape time exhibits
non-monotonic dependence on the index $\alpha$, while still retaining the
exponential form of the escape time density.
\end{abstract}


\date{\today}

\maketitle

\section{Introduction}

In his seminal paper \cite{kramers}, Kramers proposed to model chemical
reaction rates as the diffusion of a Brownian particle, initially
located in a potential well, across a potential barrier of finite height.
Meanwhile, Kramers' theory has been applied to a much more general
range of processes associated with the barrier crossing of a physical
entity experiencing random kicks fuelled by its contact to a thermal bath
\cite{pontryagin,stratonovich,haenggi,melnikov}.

Since Kramers' solution numerous different ways have been reported to access
the barrier crossing problem, in particular, to find the mathematically most
convenient formulation, compare, for instance,
Refs. \cite{stratonovich,haenggi,melnikov,chandrasekhar}. The result for the
characteristic escape time reads \cite{risken}
\begin{equation}
\label{eq4}
T_{\mathrm{esc}} = \frac{2\pi
\exp\left(\left[U(x_{\mathrm{max}})-U(x_{\mathrm{min}})\right]/D\right)
}{\sqrt{U''\left(x_{\mathrm{min}}\right)|U''\left(x_{\mathrm{max}}\right)}|}
\end{equation}
where $U\left(x\right)$ is the dimensionless potential, a prime
stands for the derivative with respect to the coordinate $x$;
$x_{\mathrm{min}}$ and $x_{\mathrm{max}}$ are the points, where the potential
$U\left(x\right)$ attains its minimum and maximum, respectively; and $D$ is
the diffusion coefficient (noise intensity) of the diffusing particle,
stemming from its coupling to the heat bath. Eq.~(\ref{eq4}) is based on the
assumption that the barrier is high (or equivalently, the noise intensity is
low), and there exists a constant probability flux across the barrier maximum.

Random processes in complex systems frequently violate the rules of Brownian
motion. Thus, the presence of static or dynamic disorder might give rise to
memory effects causing subdiffusion \cite{bouchaud,report1}, and possible
deviations from standard exponential relaxation \cite{shlekla,vlad}.
The barrier crossing in the presence of long-range memory effects was, inter
alia, modeled in terms of a generalized Langevin equation \cite{grote,haenggi1},
or via a subdiffusive fractional kinetic approach with Mittag-Leffler survival
\cite{cpl,report}. Subdiffusion is usually associated with a long-tailed waiting
time distribution $\psi(t)\sim A_{\gamma}\tau^{\gamma}/t^{1+\gamma}$ with
$0<\gamma<1$ rendering the resulting continuous time random walk process
semi-Markovian, its hallmark being the power-law time-dependence $\langle
x^2\rangle\propto K_{\gamma}t^{\gamma}$ of the mean-squared displacement
in absence of an external potential \cite{bouchaud,report1,scher}.

While in subdiffusion the waiting time between successive jump events becomes
modified such that the mean waiting time $\int_0^{\infty}t\psi(t)dt$ diverges
and consequently no natural time scale separating microscopic and macroscopic
events exists, the distribution of the lengths of individual jumps is narrow.
The converse is true for L{\'e}vy flights: 
Here, the lengths of the jumps are distributed according to the long-tailed
jump length distribution
\begin{equation}
\label{jld}
\lambda(x)\sim\frac{A_{\alpha}\sigma^{\alpha}}{|x|^{1+\alpha}}
\end{equation}
with $0<\alpha<2$ \cite{klablushle,bouchaud,report,hughes}. Thus, the
variance $\int_{-\infty}^{\infty}x^2\lambda(x)dx$ of the jump lengths
diverges.
Such power-law forms of the jump lengths have been recognized in a wide
number of fields \cite{report1}. Prominent examples for such genuine
L{\'e}vy flights are known from noise patterns in plasma devices
\cite{plasma}, and from random walks of particles or excitations along
a fastly folding polymer, where the walker is allowed to cross the small
gap between two segments of the chain, that are close by in the embedding
space due to polymer looping \cite{looping}. In the latter case, the
exponent $\alpha$ is in fact related to the critical exponents of the
polymer chain. Further examples come from fluctuations in energy space
in small systems \cite{walter,sms}, and from paleoclimatic time series
\cite{ditlevsen}.

From a mathematical point of view the occurrence of long-tailed
distributions appears quite natural due to the
L{\'e}vy-Gnedenko generalized central limit theorem \cite{levy,gnedenko}.
Indeed, the tails of probability densities of the type of $\lambda(x)$,
Eq.~(\ref{jld}), are obtained from the characteristic function of a symmetric
$\alpha$-stable distribution of the form
\begin{equation}
\label{char}
\lambda(k)\equiv\mathscr{F}\{\lambda(x)\}=\int_{-\infty}^{\infty}e^{ikx}
\lambda(x)dx=\exp\left(-c|k|^{\alpha}\right),
\end{equation}
where $c>0$. The power-law asymptotics at large $|x|$, Eq.~(\ref{jld}), appear
immediately from the expansion of the characteristic function Eq.~(\ref{char})
in the limit of small $k$.

An important question arises when replacing the Brownian particle in a barrier
crossing process by a particle executing L{\'e}vy flights. This situation can
be modelled by a particle subject to L{\'e}vy stable noise, on the level of
the Langevin equation. First steps in this direction of addressing have been
taken, as reported in Refs.~\cite{ditlevsen1,epl,bao,imkeller}. In the present
work, we report extended simulations results for the usually studied case of
low noise strength $D$, observing a pronounced step-like behavior of the
dependence of the exponent $\mu$ on the L{\'e}vy index $\alpha$. We also
explore the case of intermediate and high noise strength, finding a quite
rich behavior in the parameter space, including an optimum $\alpha$ for
the escape time.

\section{Underlying Langevin equation}

We start from the Langevin equation for a particle embedded in an external
potential field and subjected to a random noise,
\begin{equation}
\label{langevin}
\frac{dx(t)}{dt}= -\frac{1}{m\gamma}\frac{dU(x)}{dx}+\xi_\alpha(t),
\end{equation}
where $x(t)$ is the dynamic variable (particle position), $m$ the mass,
$\gamma$ the friction constant, and $U(x)$ an external potential. The noise
$\xi_\alpha(t)$ is a white, symmetric $\alpha$-stable noise.
Eq.~(\ref{langevin}) is understood in the following way \cite{plasma1}.
Integrating Eq.~(\ref{langevin}) over the interval $[t,t+\Delta t]$, we
obtain
\begin{equation}
\label{intlangevin}
x(t+\Delta t)-x(t)=-\frac{1}{m\gamma}\int_t^{t+\Delta t}\frac{dU(x(t'))}{dx}
dt'+L_{\alpha,D}(\Delta t),
\end{equation}
where
\begin{equation}
\label{noiseint}
L_{\alpha,D}(\Delta t)=\int_t^{t+\Delta t}\xi_{\alpha,D}(t')dt'
\end{equation}
is an $\alpha$-stable process with stationary independent increments and
characteristic function
\begin{equation}
\label{chnoise}
p_L(k,\Delta t)=\exp\left(-D|k|^{\alpha}\Delta t\right).
\end{equation}
$D$ is the intensity of the L{\'e}vy noise. With the use of
Eqs.~(\ref{noiseint}) and (\ref{chnoise}) it is straightforward to show
\cite{plasma1} that the discrete time representation of Eq.~(\ref{intlangevin})
at times $t_n=n\delta t$, $n=0,1,2,\ldots$ for sufficiently small time
step $\delta t$ is
\begin{equation}
\label{dlangevin}
x_{n+1}-x_n=-\frac{1}{m\gamma}\frac{dU(x_n)}{dx}\delta t+\left(D\delta t
\right)^{1/\alpha}\xi_{\alpha,1}(n),
\end{equation}
where $\{\xi_{\alpha,1}(n)\}$ is a set of random variables possessing
L{\'e}vy stable distribution $\lambda(x)$ with the characteristic function
(\ref{char}) and $c=1$.

\section{Simulations results}

\begin{figure*}
\includegraphics[width=\textwidth]{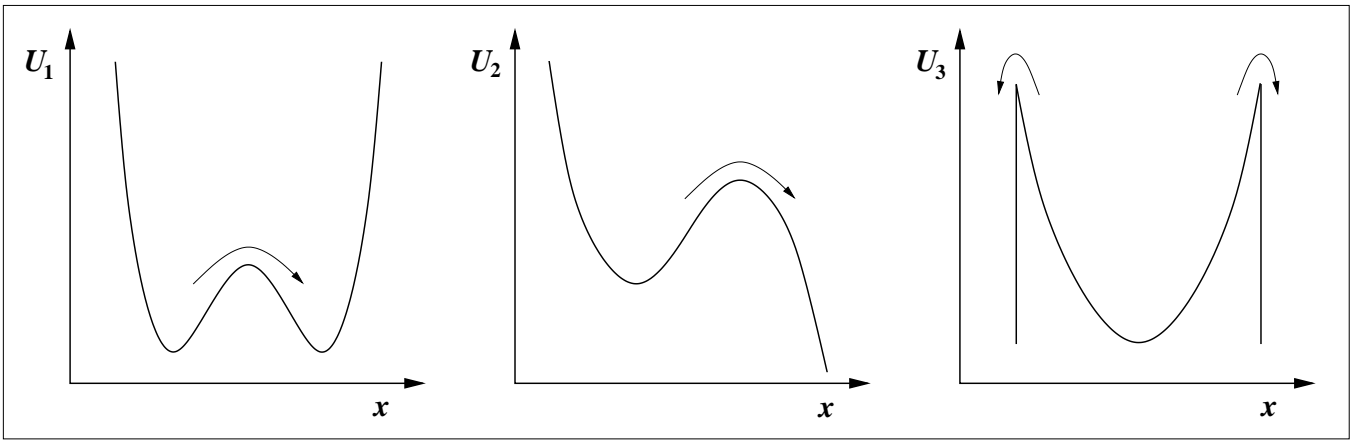}
\caption{\label{SHAPES} The three types of potential profiles $U_1$ (bistable
potential), $U_2$ (metastable potential), and $U_3$ (truncated harmonic
potential) considered in the simulations; for $g_1,\,g_2,\,L>0$.}
\end{figure*}

Before addressing the barrier crossing problem in the presence of L{\'e}vy
stable noise analytically below, we present results from extensive
simulations. In these simulations, we employ three types of potential
profiles: bistable, metastable, and truncated harmonic potentials (see
Fig.\ref{SHAPES}) defined as follows:
\begin{subequations}
\label{pots}
\begin{eqnarray}
\label{bist}
&&U_1(x)=-g_1\frac{x^2}{2}+g_2\frac{x^4}{4};\\
&&U_2(x)=-g_1\frac{x^3}{3}+{g_2}x\\
&&U_3(x)=\left\{
\begin{array}{ll}
g_1x^2/2, & -L \leq x \leq L\\[0.2cm]
0, & |x|>L\\
\end{array}  \right..
\end{eqnarray}
\end{subequations}
Let us turn to dimensionless variables. To this end we substitute $x\to x/
\tilde{x}$, $t\to t/\tilde{t}$, and $D\to D/\tilde{D}$ into the discrete time
Langevin equation (\ref{dlangevin}), and choose the appropriate constants
for rescaling, $\tilde{x}$, $\tilde{t}$, and $\tilde{D}$, for each potential
type. Taking into account that $\xi_{\alpha}\left(t/\tilde{t}\right)=
\tilde{t}^{1-1/\alpha}\xi_{\alpha}(t)$ \cite{plasma1}, we find
\begin{subequations}
\begin{eqnarray}
&&\tilde{x}=\sqrt{\frac{g_2}{g_1}},\, \tilde{t}=\frac{g_1}{m\gamma},\,
\tilde{D}=\frac{m\gamma}{g_1}\left(\frac{g_2}{g_1}\right)^{\alpha/2};\\
&&\tilde{x}=\sqrt{\frac{g_1}{g_2}},\, \tilde{t}=\frac{\sqrt{g_1g_2}}{m\gamma},
\, \tilde{D}=\frac{m\gamma}{\sqrt{g_1g_2}}\left(\frac{g_1}{g_2}\right)^{
\alpha/2};\\
&&\tilde{x}=L,\,
\frac{\tilde{x}^{\alpha}}{{\tilde{D}}}=\tilde{t}=\frac{m\gamma}{g_1},
\end{eqnarray}
\end{subequations}
respectively. In terms of these rescaled variables the discrete time Langevin
equation (\ref{dlangevin}) reads
\begin{equation}
\label{rescdlang}
x_{n+1}-x_n=-\frac{dU(x_n)}{dx}\delta t+\left(D\delta t\right)^{1/\alpha}
\xi_{\alpha,1}(n),
\end{equation}
where the potential $U(x_n)$ refers to, respectively, $U_1$, $U_2$, or $U_3$
from Eqs.~(\ref{pots}), with $g_1=g_2=L=1$.

In the simulations, for each type of the potential the particle starts from
the bottom of the potential well. The adsorbing boundary is located: (i) for
the bistable potential in the saddle point $x=0$; (ii) for the metastable
potential far to the right of the saddle point, at $x=10$; and (iii) for the
truncated harmonic potential at the boundaries, $|x|=1$. The process defined
by Eq.~(\ref{rescdlang}) is repeated until the particle reaches, or crosses
the adsorbing boundary; then the process in Eq.~(\ref{rescdlang}) is restarted.
This procedure is performed 100,000 times, for fixed $D$ and $\alpha$, and the
mean escape time is then calculated \cite{REMM}.

\begin{figure}
\includegraphics[width=8.6cm]{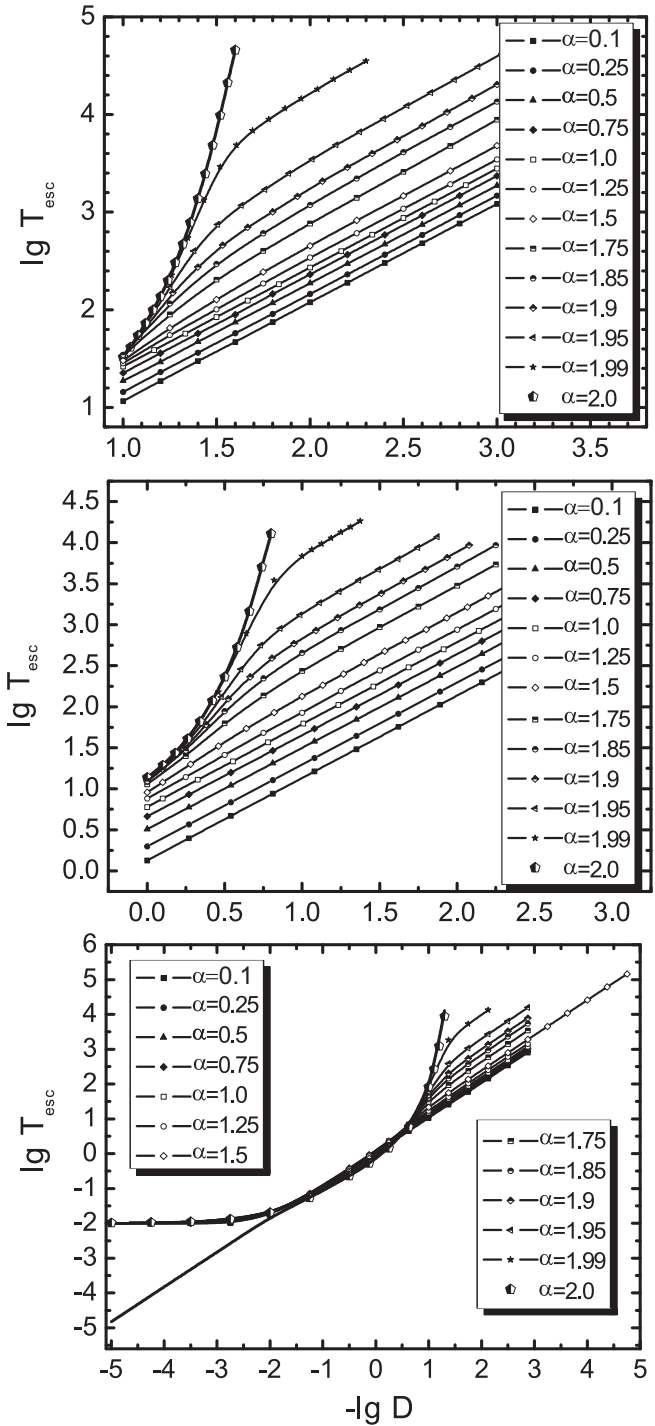}
\caption{\label{MET} Mean escape times as function of noise
intensity $D$ for the bistable (top), metastable (middle), and truncated
harmonic potentials (bottom).}
\end{figure}

The mean escape times $T_{\mathrm{esc}}$ of the diffusing particle as function
of noise intensity $D$ for the three potential profiles are shown in
Fig.~\ref{MET} for different values of $\alpha$ ranging from 0.1 to 2
\cite{REMM1}. It is clear that the curves for $\alpha<2$ obey a different law
in comparison to their Gaussian counterpart. We observe three different regimes
for the dependence of $T_{\mathrm{esc}}$ on the L{\'e}vy noise parameters,
which can be classified by the value of noise intensity. Let us discuss these
regimes in detail.

\subsection{Low noise intensity regime}

\begin{figure}
\includegraphics[width=8.6cm]{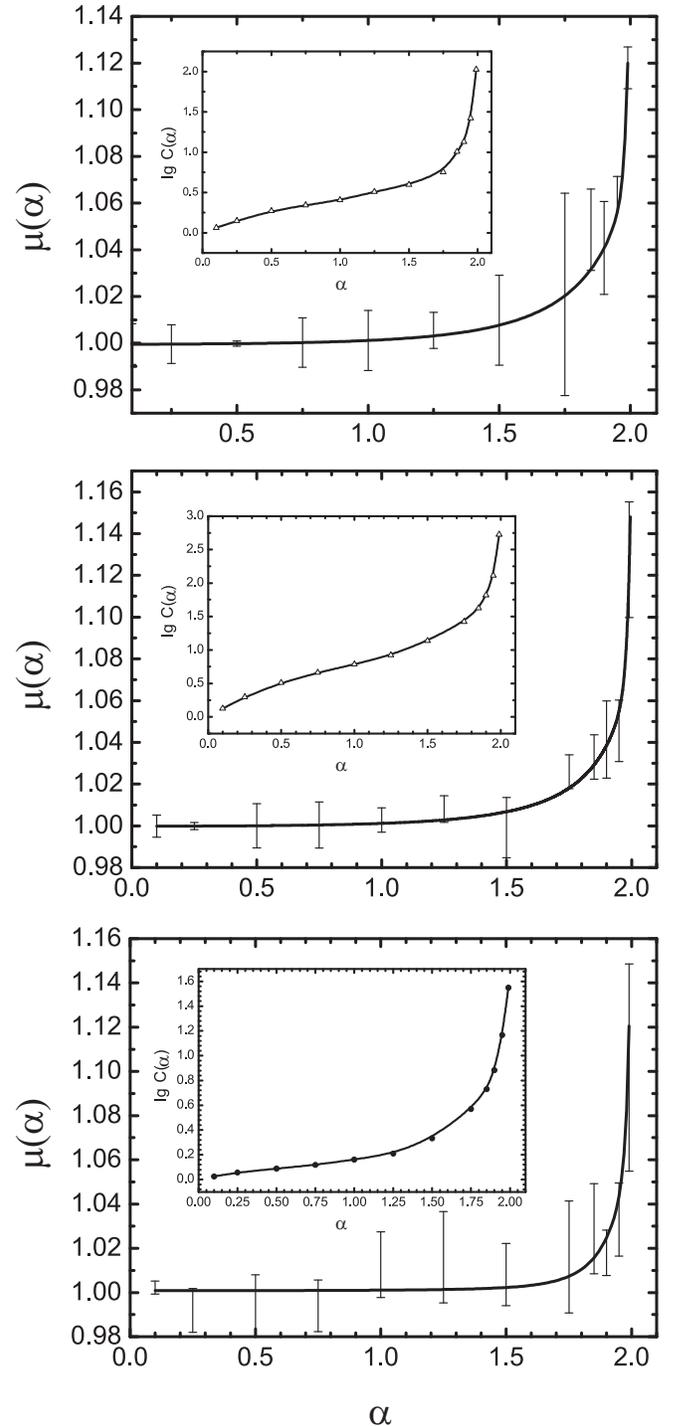}
\caption{\label{MU}
Dependencies of $\mu(\alpha)$ and $C(\alpha)$ for the bistable (top),
metastable (middle) and truncated harmonic (bottom) potentials. See text.}
\end{figure}

In this regime, instead of an exponential dependence on $1/D$, the curves
shown in Fig.~\ref{MET} display a power-law asymptotic behavior of the mean
escape time
\begin{equation}
T_{\mathrm{esc}}(\alpha,D)=\frac{C(\alpha)}{D^{\mu(\alpha)}}.
\end{equation}
Further analysis of the data plotted in Fig.~\ref{MET} allows us to determine
the dependencies of the power law exponent $\mu(\alpha)$ and coefficient
$C(\alpha)$ on the L{\'e}vy index $\alpha$ (see Fig.~\ref{MU}). The exponent
$\mu(\alpha)$ is approximately unity up to $\alpha\approx 1.5$. The dependence
of $C(\alpha)$ for the first two potential types possesses a weak inflection
at small $\alpha$. Note that for fixed noise strength $D$, the dependence of
$T_{\mathrm{esc}}$ on the L{\'e}vy index $\alpha$ is monotonic.

\begin{figure}
\includegraphics[width=8.6cm]{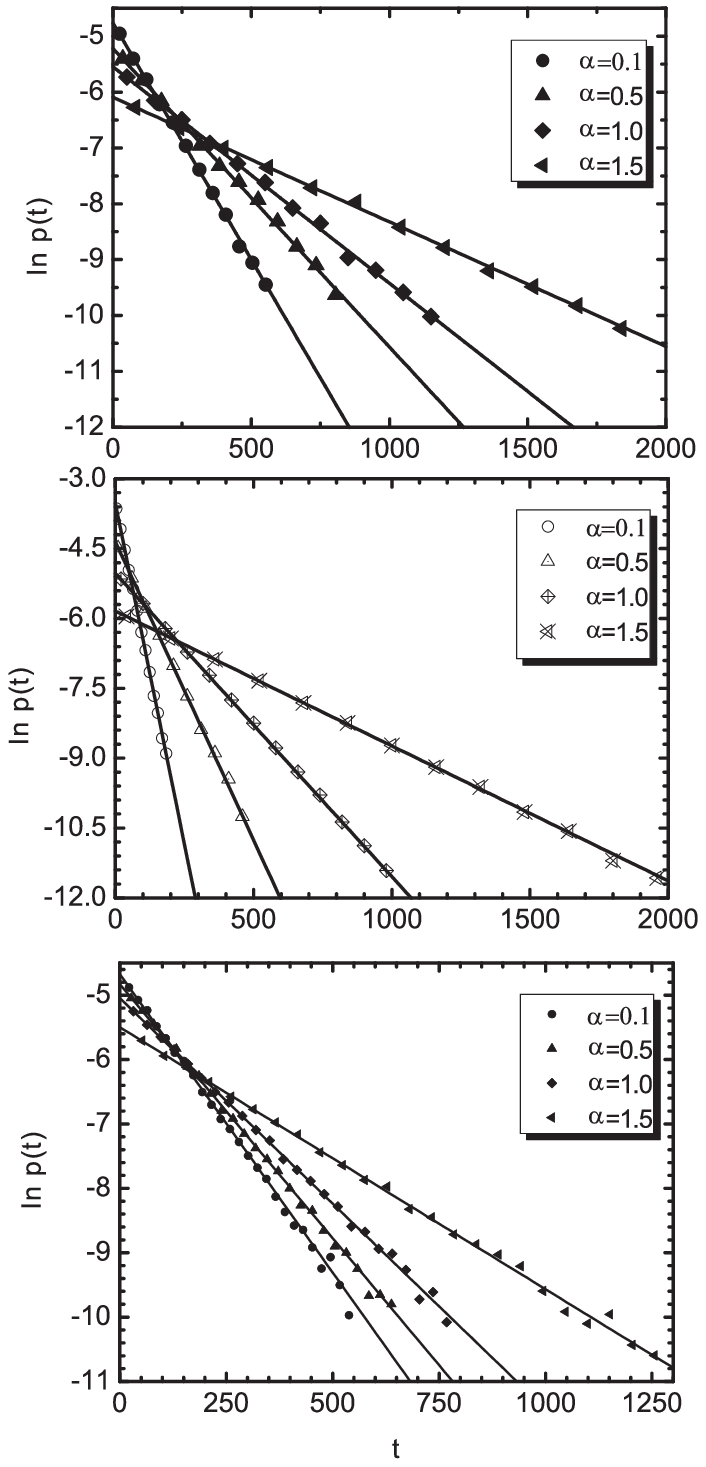}
\caption{\label{FigPDF}
Escape time probability density functions in dependence of time
for the bistable (top), metastable (middle) and truncated harmonic
(bottom) potentials. In the log versus lin plot, the exponential
dependence is obvious.}
\end{figure}

Let us now turn to the probability density function (PDF) $p(t)$ of the escape
times. We use again our simulation scheme to obtain $p(t)$ from the Langevin
equation (\ref{rescdlang}), but now for each fixed values of $\alpha$ and $D$
we collect 200,000 escape events, which are not averaged but processed with a
simple routine, that constructs the PDF. The results for the three potentials
from Fig.\ref{SHAPES} are shown in Fig.~\ref{FigPDF} \cite{REMM2}. It is
imminently clear that in all three cases and for all $\alpha$ (including the
Gaussian case, $\alpha=2$) the exponential decay pattern
\begin{equation}
\label{surv}
p(t)=\frac{1}{T}\exp\left(-\frac{t}{T}\right)
\end{equation}
is nicely obeyed, in agreement with the findings from previous simulations
studies \cite{ditlevsen,epl,bao}, as well as with the analytical results
reported in Ref.~\cite{imkeller}.

Relation (\ref{surv}) allows us to extract the mean escape time $T_{\mathrm{
esc}}$ independently from our previous results in Fig.~\ref{MET}, in two
different ways. Namely, we define
\begin{equation}
\label{t1}
T_{\mathrm{esc}}=\frac{1}{p(0)}\equiv T_1;
\end{equation}
alternatively, we determine
\begin{equation}
\label{t2}
T_{\mathrm{esc}}=-\left(\frac{d\ln p(t)}{dt}\right)^{-1}
\equiv T_2.
\end{equation}
These two ways to determine $T_{\mathrm{esc}}$ are intrinsically different,
$T_1$ depending on the extrapolation of the fitted exponential behavior
to $t=0$, whereas $T_2$ is determined through the slope in the logarithmic
versus linear plot. Tab.~\ref{Tbl1} shows the results for $T_1$ and $T_2$
along with the values of $T_{\mathrm{esc}}$ obtained directly from the
simulations results in Figs.~\ref{MET}. Indeed, the value of $T_2$ is
consistently closer to $T_{\mathrm{esc}}$ than $T_1$. Overall, however, the
agreement is very good (better than 1.5 per cent).

\begin{table}
\begin{ruledtabular}
\begin{tabular}{cccc}
\multicolumn{4}{c}{First potential type, $D=10^{-2.0}$ \hfill} \\
\hline
$\alpha$ & $T_{\mathrm{esc}}$ & $T_1$ & $T_2$ \\
\hline
0.1 & 119.7 & 117.4 & 118.1 \\
0.5 & 187.1 & 185.1 & 187.1 \\
1.0 & 260.8 & 257.1 & 258.2 \\
1.5 & 446.5 & 443.4 & 448.4 \\
\hline
\multicolumn{4}{c}{Second potential type, $D=10^{-1.4}$ \hfill} \\
\hline
$\alpha$ & $T_{\mathrm{esc}}$ & $T_1$ & $T_2$ \\
\hline
0.1 & 34.2 & 33.9 & 34.1 \\
0.5 & 78.3 & 77.8 & 78.1 \\
1.0 & 153.7 & 153.0 & 153.5 \\
1.5 & 346.6 & 342.9 & 344.9 \\
\hline
\multicolumn{4}{c}{Third potential type, $D=10^{-2.0}$ \hfill} \\
\hline
$\alpha$ & $T_{\mathrm{esc}}$ & $T_1$ & $T_2$ \\
\hline
0.1 & 108.2 & 107.1 & 107.9 \\
0.5 & 127.4 & 125.4 & 126.8 \\
1.0 & 159.1 & 155.7 & 156.7 \\
1.5 & 250.2 & 244.6 & 245.9 \\
\end{tabular}
\end{ruledtabular}
\caption{\label{Tbl1}
Mean escape times, obtained using three separate methods: $T_{\mathrm{esc}}$
is determined directly from the simulations producing Fig.~\ref{MET}, while
$T_1$ and $T_2$ are defined in Eqs.~(\ref{t1}) and (\ref{t2}), respectively.}
\end{table}

\subsection{High and intermediate noise intensity regimes}

\begin{figure}
\includegraphics[width=8.6cm]{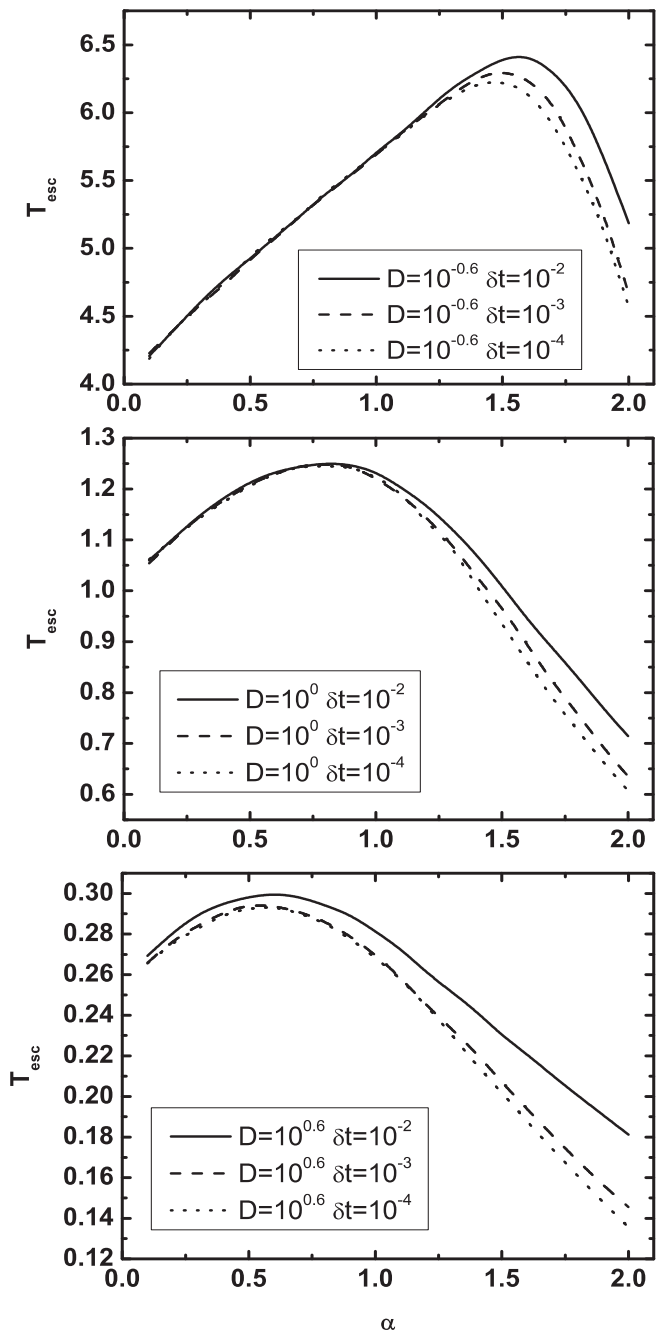}
\caption{Dependence of mean escape time $T_{\mathrm{esc}}$ on L{\'e}vy
index $\alpha$, for various noise intensities $D$ in the intermediate
regime. For these results we used the truncated harmonic potential.
The weak dependence on the time increment $\delta t$ of the simulation
is discussed in the text.}
\label{fig5}
\end{figure}

\begin{figure}
\includegraphics[width=8.6cm]{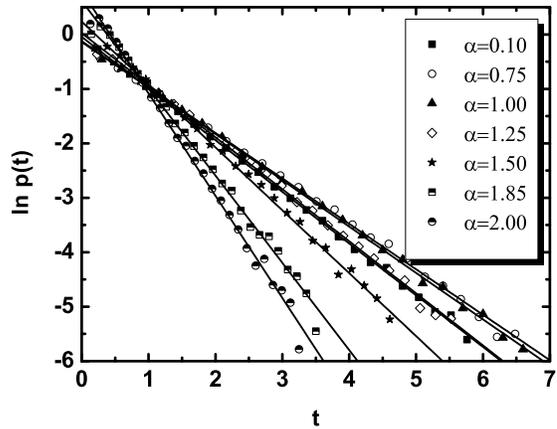}
\caption{Escape time PDF for the intermediate noise strength, with
$D=1$. Note that the curves for $\alpha=0.1$ and 1.25, as well
as $\alpha=0.75$ and $1.0$ are almost coincide, pointing at a weak
inversion of the $\alpha$-dependence.}
\label{nf}
\end{figure}

We consider these two regimes for the case of the truncated harmonic potential
as example, compare to the bottom graph in Fig.\ref{MET}. The two other cases,
bistable and metastable potentials, demonstrate similar behavior. When the
noise intensity $D$ increases beyond the low noise limit, in which the curves
exhibit a linear dependence and are almost parallel, the various curves for
the escape time show a tendency to collapse. This is the intermediate noise
intensity
regime, which for the truncated harmonic potential lies approximately in the
range $-2<-lg D<1$. Here, we observe that $T_{\mathrm{esc}}$ demonstrates very
weak dependence on the L{\'e}vy index, in comparison with the low intensity
regime. The explanation is the following. Contrary to the weak noise intensity
regime, where the tail of the L{\'e}vy noise distribution plays a decisive role in
the escape process (less frequent but high amplitude noise spikes, that is,
govern the barrier crossing), here the main role in the process, due to the
higher intensities, passes on to the central part of the distribution. Since
all stable distributions, including the Gaussian, have central parts which are
of similar shape, the $\alpha$-dependence of the escape time becomes weaker.
Furthermore, in the intermediate region we observe a non-monotonic dependence
of $T_{\mathrm{esc}}$ versus $\alpha$, examples for three values of $D$ taken
from the intermediate regime are shown in Fig.~\ref{fig5}. This weak
non-monotonicity most likely reflects minor differences in the shapes of the
distribution of the noise $\xi_{\alpha}$ in the central part (relative to the
much more pronounced differences in the tails of these PDFs). This point will
be investigated in more detail in a forthcoming paper.
Another important feature of the intermediate regime is that the PDF of the
escape times remains exponential. This is clearly seen in Fig.~\ref{nf}. In
that figure, one can also recognize that for $\alpha=0.1$ and 1.25, as well
as $\alpha=0.75$ and $1.0$ are almost coinciding.

In Fig.~\ref{fig5} we also show the dependence of the mean escape time 
on the time increment $\delta t$ used in our simulations, showing a much
higher sensitivity to the particular value chosen for $\delta t$ than
in the regime of low noise strength. This is intuitively clear, as now
the process is terminated after few jumps only. As seen in Fig.~\ref{fig5},
for increasingly small $\delta t$, the shown curves converge. Note, however,
that the weak non-monotonicity of the escape time on $\alpha$ is preserved for
all $\delta t$.

When the noise intensity $D$ is increased further, the intermediate regime
turns over to the high noise intensity regime. Here, all the curves for the
escape time $T_{\mathrm{esc}}$ collapse into a single curve, that approaches
the constant value $T_{\mathrm{esc}}=10^{-2}$, independent of $D$. Noting that
in our simulation scheme the time step $\delta t=10^{-2}$ is used, we conclude
that in the region of high noise intensity the particles reach the boundary in
a single jump. We confirmed this conclusion by taking different values of the
time step, observing that, whereas the picture in the low and intermediate
regions changes negligibly, in the high intensity regime the collapsed curves
again approach the value of $T_{\mathrm{esc}}$ equal to a single time step.

\section{Analytical results for the Cauchy case, $\alpha=1$}

A standard approach to the classical barrier crossing problem in the presence
of Gaussian white noise is based on the stationary flux approximation assuming
that the probability current across the barrier is constant. This is equivalent
to requiring that the barrier is high in comparison to thermal energy, or the
noise intensity $D$ low. The stationary flux approximation is widely used for
the classical Kramers problem
\cite{kramers,stratonovich,chandrasekhar,klimontovich,risken}.
Here, we extend this assumption to the case of white L{\'e}vy noise, and
obtain analytical results for the Cauchy case, $\alpha=1$, in a bistable
potential.

We start from the fractional Fokker-Planck equation (FFPE)
\begin{equation}
\label{ffpe}
\frac{\partial f}{\partial t}=\frac{\partial}{\partial x}\left(\frac{dU}{dx}
\frac{f}{m\gamma}\right)+D\frac{\partial^\alpha f}{\partial |x|^\alpha},
\end{equation}
for the density $f(x,t)$ to find the diffusing particle at position $x$ at
time $t$ \cite{report}. Here, the potential $U(x)$ is defined in
Eq.~(\ref{bist}), and the fractional Riesz derivative $\partial^{\alpha}/
\partial|x|^{\alpha}$ is understood via its Fourier transform,
\begin{equation}
\mathscr{F}\left\{\frac{\partial^{\alpha}}{\partial|x|^{\alpha}}f(x,t)\right\}
\equiv\int_{-\infty}^{\infty}e^{ikx}\frac{\partial^{\alpha}}{\partial|x|^{
\alpha}}f(x,t)dx=-|k|^{\alpha}f(k,t).
\end{equation}
We note that the FFPE (\ref{ffpe}) is equivalent to the Langevin approach in
Eqs.~(\ref{langevin}) to (\ref{chnoise}).

After rescaling in the same way as outlined above, we arrive at the FFPE
with dimensionless variables,
\begin{subequations}
\begin{equation}
\label{eqn3}
\frac{\partial f}{\partial t}=\frac{\partial}{\partial x}\left(\frac{dU}{dx}f
\right)+D\frac{\partial^{\alpha}f}{\partial |x|^\alpha},
\end{equation}
where
\begin{equation}
U(x)=-\frac{x^2}{2}+\frac{x^4}{4}.
\end{equation}
\end{subequations}
We can rewrite Eq.~(\ref{eqn3}) in terms of the flux $j$,
\begin{equation}
\label{fluxy}
\frac{\partial f}{\partial t}+\frac{\partial j}{\partial x}=0.
\end{equation}
We follow the stationary flux approximation of the Brownian theory, and also
pursue the assumptions used to obtain the escape time in the Brownian limit
\cite{klimontovich}, and can be justified in a more stringent approach
\cite{stratonovich}. Namely, we solve the stationary Eq.~(\ref{ffpe}), under
the assumption that the stationary distribution of particles differs
significantly from the equilibrium distribution: the ``source'' ensures that
all particles are on the left-hand side of the barrier, causing the steady
influx of particles toward the barrier; after crossing it, the particles
disappear through the `sink'”. That is, following Ref.~\cite{klimontovich} we
assume
\begin{subequations}
\begin{equation}
\label{normal}
\int_{-\infty}^0f(x)dx=1,
\end{equation}
and
\begin{equation}
\label{conflu}
j(x)=j_0=T_{\mathrm{esc}}^{-1},
\end{equation}
\end{subequations}
where $j_0$ is the constant flux of particles across the barrier, and $T_{
\mathrm{esc}}$ corresponds to the mean first passage time.

Transforming to Fourier space, the FFPE becomes
\begin{equation}
\label{fffpe}
\frac{\partial f}{\partial t}=k\frac{\partial f}{\partial k}+k\frac{\partial^3
f}{\partial k^3}-D|k|^{\alpha}f(k),
\end{equation}
while Eq.~(\ref{fluxy}) attains the form
\begin{equation}
\label{eqn7}
\frac{\partial f}{\partial t}-ikj(k)=0.
\end{equation}
Comparing Eqs.~(\ref{fffpe}) and (\ref{eqn7}), we obtain for the
Fourier transform of the flux the expression
\begin{equation}
\label{eqn8}
j(k)=-i\frac{\partial^3f}{\partial k^3}-i\frac{\partial f}{
\partial k}+iD\mathrm{sign}(k)|k|^{\alpha-1}f.
\end{equation}
Now, with the use of Eq.~(\ref{conflu}) it follows from Eq.~(\ref{eqn8}) that
the stationary solution in the constant flux approximation is determined by
the following equation \cite{REMM3}:
\begin{equation}
\label{eqn25}
\frac{d^3f}{dk^3}+\frac{df}{dk}-D\mathrm{sign}(k)f=
2\pi ij_0\delta(k).
\end{equation}

We solve Eq.~(\ref{eqn25}) on the right and left semi-axes, and then
match the solutions. The details are presented in Appendix \ref{match}.
This procedure yields
\begin{equation}
\label{yield}
f(k)=\left\{\begin{array}{ll}j_0\frac{\pi}{2ab}\exp\left(z^*k\right), &
k\ge0,\\[0.2cm]
j_0\frac{\pi}{2ab}\exp\left(-zk\right), & k<0\end{array}\right.,
\end{equation}
where the constants $a$ and $b$, and $z$ are given in Eqs.~(\ref{eqn51a}).
The inverse Fourier transform is indeed a PDF, namely
\begin{eqnarray}
\nonumber
f(x)&=&\int_{-\infty}^{\infty}e^{-ikx}f(k)\frac{dk}{2\pi}\\
\nonumber
&=&\frac{j_0}{2ab}\Re\int_0^{\infty}e^{k(ix+z)}dk\\
&=&\frac{j_0}{2b}\frac{1}{(x+b)^2+a^2}.
\label{cauch}
\end{eqnarray}

The last step is the normalization conditions given by Eq.~(\ref{normal}).
After integrating Eq.~(\ref{cauch}),
\begin{equation}
\frac{j_0}{2ab}\left(\frac{\pi}{2}+\arctan\left(\frac{b}{a}\right)\right)=1,
\end{equation}
we arrive at the final expression for the mean escape time,
\begin{equation}
T_{\mathrm{esc}}=\frac{\pi}{4ab}\left(1+\frac{2}{\pi}\arctan\left(\frac{b}{a}
\right)\right).
\end{equation}
For $D\ll1$, we obtain from Eqs.~(\ref{limit1}) and (\ref{limit2}) the
asymptotic result for the escape time
\begin{equation}
\label{escape}
T_{\mathrm{esc}}\approx\frac{\pi}{D},\,\,D\ll1.
\end{equation}
This result agrees with numerical simulations using the L{\'e}vy noise
$\xi_{\alpha}$ as stochastic force, within the accuracy of about 12 per cent.

\section{Conclusions}

We consider symmetric L{\'e}vy flights within the entire domain of L{\'e}vy
indices, $0<\alpha<2$, in three generic types of the potential wells: bistable
potential, metastable potential, and truncated harmonic potential. As the
basis for the numerical analysis we use the Langevin equation with the L{\'e}vy
noise as source for the stochastic force, whereas for the analytical treatment
we employ the space fractional Fokker-Planck equation. We obtain the following
results.

Firstly, we demonstrated by extensive numerical analysis based on solving
the Langevin equation the existence of \emph{three\/} dynamic regimes in the
barrier crossing behavior of a particle driven by L{\'e}vy noise, namely:

(i) The regime of low noise intensity $D$, displaying an algebraic dependence
of the mean escape time on $D$ through $T_{\mathrm{esc}}\simeq C(\alpha)/D^{\mu
(\alpha)}$. In this regime the nature of the L{\'e}vy noise with its
characteristic property to allow for large outliers, plays an essential role
and dominates the escape process. We showed that the exponent $\mu(\alpha)$
remains approximately constant, $\mu\approx1$ for $0<\alpha<2$; close to
$\alpha=2$, it displays a divergence towards the exponential dependence on
$1/D$ at $\alpha=2$, the case of Gaussian noise. In this low noise intensity
regime, we observe a monotonous increase of the escape time with increasing
$\alpha$ (keeping the noise intensity $D$ fixed). The probability density of
escape times decays exponentially.

(ii) The regime of intermediate noise intensity, in which the escape time is
determined by the central part of probability distribution of the noise. This
regime is characterized by the following features: The difference between the
escape times for different L{\'e}vy indices significantly decreases, since in
the central part all stable distributions are very similar to each other, as
well as to the Gaussian distribution. Additionally, there is a non-monotonic
dependence of the escape time with increasing $\alpha$. However, the
probability densities of escape times still decay exponentially.

(iii) The universal regime of high noise intensity, where the particle
escapes with the first step (or in very few steps). The curves denoting the
dependence of the escape time $T_{\mathrm{esc}}$ on noise intensity $D$
collapse onto a single curve for all values of $\alpha$ (including the
Gaussian limit).

Secondly, for the particular Cauchy case, $\alpha=1$ we develop the kinetic
theory of the escape over the barrier in a bistable potential. We start from the
space fractional Fokker-Planck equation and use the assumption of a constant
flux over the barrier. We find analytically the expression for the escape time,
which at low noise intensities produces the result $T_{\mathrm{esc}}\approx
\pi/D$. This result agrees with numerical simulation within the accuracy of
about 12 per cent.

\acknowledgments

We wish to thank V.~Yu.~Gonchar and T.~Koren for their help with the
simulations, and M.~Lomholt for discussions.
RM acknowledges partial funding through the Natural Sciences and Engineering
Research Council (NSERC) of Canada, and the Canada Research Chairs programme.

\begin{appendix}

\section{Simulations description}
\label{simulations}

In this Appendix, we briefly outline our simulations strategy, in particular,
how we simulate the white L{\'e}vy noise $\xi_\alpha\left(n\right)$. The
corresponding generator is taken from \cite{chambers}. We start by calculating
the value
\begin{equation}
X=\frac{\sin\left(\alpha\gamma\right)}{\left(\cos\gamma\right)^{1/\alpha}}
\left(\frac{\cos\left([1-\alpha]\gamma\right)}{W}\right)^{\left(1-\alpha\right)
/\alpha},
\end{equation}
where $\gamma$ is a random value uniformly distributed on the interval
$\left(-\pi/2,\pi/2\right)$, $W$ is an independent random variable with
exponential distribution, and $\alpha$ is the L{\'e}vy index, $0<\alpha\le2$.

The proof that the calculated variable $X$ possesses a L{\'e}vy distribution
function goes as follows. Firstly, let $0<\alpha<1$. While $\gamma>0$,
the expression for $X$ may be presented as
\begin{equation}
X=\left(\frac{a(\gamma)}{W}\right)^{\left(1-\alpha\right)/\alpha},
\end{equation}
where
\begin{equation}
a\left(\gamma\right)=\left(\frac{\sin\alpha\gamma}{\cos\gamma}\right)^{\alpha/
(1-\alpha)}\cos\left([1-\alpha]\gamma\right).
\end{equation}
Then
\begin{eqnarray}
\nonumber
P\left(0\le X\le x\right)&=&P\left(0\le X\le x,\gamma>0\right)\\
\nonumber
&&\hspace*{-1.2cm}=P\left(0\le\left(\frac{a(\gamma)}{W}\right)^{1/\alpha-1}
\le x,\gamma>0\right)\\
\nonumber
&&\hspace*{-1.2cm}=P\left(W\ge x^{-\alpha/(1-\alpha)}a(\gamma),\gamma>0
\right)\\
\nonumber
&&\hspace*{-1.2cm}\equiv\frac{1}{\pi}\int_0^{\pi/2}d\gamma\int_{a(\gamma)
x^{\alpha/(1-\alpha)}}^{\infty}{d\xi}e^{-\xi}\\
&&\hspace*{-1.2cm}=\int_0^{\pi/2}d\gamma\exp{\left(-x^{-\alpha/(1-\alpha)}
a\left(\gamma\right)\right)}.
\end{eqnarray}
The latter expression, according to Ref.~\cite{zolotarev} is an integral
representation of L\'evy distribution.

In the case $1<\alpha\le 2$ analogous steps pertain, but now for the quantity
$P\left(X\ge x,\gamma>0\right)$. When $\alpha=1$, the relation for $X$ turns
into $X=\tan \gamma$, that, is known to be a random value with Cauchy PDF. One
does not need to consider the case $\gamma<0$: it merely corresponds to
negative $X$ values.

The numerical simulation of $T_{\mathrm{esc}}$ as function of $D$ is performed
according to the following flowchart:
\begin{enumerate}
\item A ``particle'' is placed at the potential minimum;
\item A value of $\alpha$ is fixed;
\item The discrete-time Langevin equation (\ref{dlangevin}) is iterated, until
the particle reaches a definite coordinate, namely $x=0$ for the bistable
potential, $x=10$ for the metastable potential, and $x=\pm 1$ for the 
truncated harmonic potential;
\item The event time $t=n\Delta t$ is stored;
\item Steps 3 and 4 are executed 10,000 times for each fixed
value of $D$, and the average escape time calculated;
\end{enumerate}

A similar procedure is used to simulate the escape time PDF $p(t)$, apart
from taking the average. Explicitly:
\begin{enumerate}
\item The ``particle'' is placed at the potential minimum;
\item Some value of $D$ is fixed;
\item Some value of $\alpha$ is taken;
\item The discrete-time Langevin equation (\ref{dlangevin}) is iterated,
until the ``particle'' reaches a definite coordinate, namely $x=0$ for the
bistable potential, $x=10$ for the metastable potential, and $x=\pm1$ for
the truncated harmonic potential;
\item The event time $t=n \Delta t$ is stored;
\item Step 4 is executed 200,000 times for each fixed
value of $\alpha$, but, this time, the obtained event times are
not averaged, but handled with a simple routine, that builds the
PDF;
\end{enumerate}

Each run was repeated and both results compared. Throughout, high
reliability was observed.

\section{Matching procedure}
\label{match}

In this Appendix, we collect the necessary steps to identify the constants
for the determination of the density $f$ in the barrier crossing for the
Cauchy case.

To this end, we solve Eq.~(\ref{eqn25}) on the right and left semi-axes,
and then match the solutions. For the two domains, we make the exponential
ansatz
\begin{equation}
\label{eqn26}
f(k)=\left\{
\begin{array}{ll}
C_1e^{zk}+C_2e^{z^*k},\,\, k>0;\\
C_3e^{\zeta k}+C_4e^{\zeta^*k},\,\,k<0,
\end{array}
\right.
\end{equation}
where
\begin{equation}
z=-\frac{u_++v_+}{2}+i\frac{u_+-v_+}{2}\sqrt{3},
\end{equation}
with
\begin{subequations}
\label{eqn30}
\begin{eqnarray}
&&u_+^3=\frac{D}{2}\left[1+\sqrt{1+\frac{4}{27D^2}}\right],\\
&&v_+^3=\frac{D}{2}\left[1-\sqrt{1+\frac{4}{27D^2}}\right].
\end{eqnarray}
\end{subequations}
Moreover, we define
\begin{equation}
\zeta=-\frac{u_-+v_-}{2}+i\sqrt{3}\frac{u_--v_-}{2},
\end{equation}
with
\begin{subequations}
\begin{eqnarray}
&&u_-^3=\frac{D}{2}\left[-1+\sqrt{1+\frac{4}{27D^2}}\right]=-v_+^3,\\
&&v_-^3=\frac{D}{2}\left[-1-\sqrt{1+\frac{4}{27D^2}}\right]=-u_+^3,
\end{eqnarray}
\end{subequations}
which implies that
\begin{equation}
\label{eqn36a} \zeta  = \frac{{u_ +   + v_ +  }} {2} + \frac{{u_ + -
v_ + }} {2}i\sqrt 3 \;,\quad z^ *   =  - \zeta.
\end{equation}

To determine the unknown complex constants $C_1, C_2, C_3, C_4$ we
match the results at $k=0$. Thus, firstly, we require $f(k-0)=f(k
+0)$, which, using Eq.~(\ref{eqn26}) yields
\begin{equation}
\label{eqn38}
C_1+C_2=C_3+C_4.
\end{equation}

The second condition is obtained by integration of Eq.~(\ref{eqn25})
over the small region $[-\varepsilon,\varepsilon]$:
\begin{equation}
\int_{-\varepsilon}^{\varepsilon}dkf'''+\int_{-\varepsilon}^{\varepsilon}
dkf'-D\int_{-\varepsilon}^{\varepsilon}dk\mathrm{sign}(k)f=2\pi
ij_0.
\end{equation}
In the limit $\varepsilon \rightarrow 0$ we recover the condition
\begin{equation}
\label{eqn40}
f''(0+)-f''(0-)=2\pi ij_0,
\end{equation}
or, after inserting Eq.~(\ref{eqn26}) into Eq.~(\ref{eqn40}),
\begin{equation}
\label{eqn41}
C_1z^2+C_2 z^{*2}-C_3\zeta^2-C_4\zeta^{*2}=2\pi ij_0.
\end{equation}
The third condition is that the PDF is a real function:
\begin{equation}
\label{eqn42}
f(x)=\int_0^{\infty}\frac{dk}{2\pi}e^{-ikx}f_1(k)+\int_{-\infty }^0
\frac{dk}{2\pi}e^{-ikx}f_2(k),
\end{equation}
and
\begin{equation}
\label{eqn43}
f^*(x)=\int_0^\infty\frac{dk}{2\pi}e^{ikx}f_1^*(k)+\int_{-\infty}^0
\frac{dk}{2\pi}e^{ikx}f_2^*(k).
\end{equation}
On the other hand, by changing $k\rightarrow-k$ in Eq.~(\ref{eqn42}),
we find
\begin{equation}
\label{eqn44}
f(x)=\int_{-\infty}^0\frac{dk}{2\pi}e^{ikx}f_1(-k)+\int_0^\infty
\frac{dk}{2\pi}e^{ikx}f_2(-k).
\end{equation}
Since $f(x)$ is a real function, we see from Eqs.~(\ref{eqn43})
and (\ref{eqn44}):
\begin{equation}
\label{eqn45}
f_1^*(k)=f_2(-k),\,\,f_2^*(k)=f_1(-k).
\end{equation}

Let us now insert Eq.~(\ref{eqn26}) into Eqs.~(\ref{eqn45}). From the
first equation we have
\begin{equation}
C_1^*e^{z^*k}+C_2^*e^{zk}=C_3e^{-\zeta k}+C_4e^{-\zeta^*k}.
\end{equation}
The second equation gives rise to the same result.
Since $z^*=-\zeta$ (see Eq.~(\ref{eqn36a})), we have
\begin{equation}
\label{eqn47}
C_1^*=C_3,\,\,C_2^*=C_4.
\end{equation}
Moreover, from Eqs.~(\ref{eqn47}) and (\ref{eqn38}) we see that
\begin{equation}
\label{eqn48}
\left(C_1+C_2\right)^*=C_3+C_4=C_1+C_2;
\end{equation}
thus, from Eqs.~(\ref{eqn47}) and (\ref{eqn48}), we obtain the following
relations:
\begin{subequations}
\begin{eqnarray}
&&C_{1R}=C_{3R},\,\,C_{2R}=C_{4R},\\
&&C_{3I}=-C_{1I},\,\,C_{4I}=-C_{2I},\\
&&C_{2I}=-C_{1I},\,\,C_{4I}=-C_{3I},\\
&&C_{2I}=C_{3I}=-C_{1I}=-C_{4I}.
\end{eqnarray}
\end{subequations}
Therefore, we are actually dealing with 3 constants only, namely, $C_{1R}$,
$C_{2R}$ and $C_I$:
\begin{subequations}
\label{eqn50}
\begin{eqnarray}
&&C_1=C_{1R}+iC_I,\,\,C_2=C_{2R}-iC_I,\\
&&C_3=C_{1R}-iC_I,\,\,C_4=C_{2R}+iC_I.
\end{eqnarray}
\end{subequations}
Inserting Eqs.~(\ref{eqn50}) into Eq.~(\ref{eqn41}) produces
\begin{eqnarray}
\nonumber
&&C_{1R}\left(z^2-\zeta^2\right)+C_{2R}\left(z^{*2}-\zeta^{*2}\right)\\
&&\hspace*{0.8cm}+iC_I\left(z^2-z^{*2}+\zeta^2-\zeta^{*2}\right)=2\pi ij_0.
\label{eqn51}
\end{eqnarray}
For convenience, we define
\begin{equation}
\label{eqn51a}
a=\frac{u_++v_+}{2},\,\,b=\frac{\sqrt{3}}{2}\left(u_+-v_+\right).
\end{equation}
Then
\begin{equation}
z=-a+ib,\,\,\zeta=a+ib;
\end{equation}
such that
\begin{subequations}
\label{eqn53}
\begin{eqnarray}
&&z^2-\zeta^2=-4iab,\\
&&z^{*2}-\zeta^{*2}=4iab,\\
&&z^2-z^{*2}+\zeta^2-\zeta^{*2}=0.
\end{eqnarray}
\end{subequations}
With the use of Eq.~(\ref{eqn53}) we get from Eq.~(\ref{eqn51}):
\begin{equation}
\label{eqn54}
C_{2R}-C_{1R}=\frac{\pi j_0}{2ab}.
\end{equation}

Now, combine Eqs.~(\ref{eqn50}) and (\ref{eqn54}), and insert into
Eqs.~(\ref{eqn26}):
\begin{widetext}
\begin{subequations}
\label{eqn55}
\begin{eqnarray}
&&f_1(k)=\left(C_{1R}+iC_I\right)e^{zk}+\left(C_{1R}+\frac{\pi j_0}{2ab}
-iC_I\right)e^{z^*k},\,\,k\ge0\\
&&f_2(k)=\left(C_{1R}-iC_I\right)e^{\zeta k}+\left(C_{1R}+\frac{\pi j_0
}{2ab}+iC_I\right)e^{\zeta^*k},\,\,k<0.
\end{eqnarray}
\end{subequations}
\end{widetext}

We are looking for the stationary (non-equilibrium) distribution; therefore we
assume $C_{1R}=C_I=0$ in order to retain in Eq.~(\ref{eqn55}) only those terms,
that are proportional to the flux $j_0$. This leads us to the result
(\ref{yield}).

To calculate the $D\ll1$ limit of the characteristic escape time
(\ref{escape}), we obtain the following limiting behaviours. Thus, from
Eqs.~(\ref{eqn30}) and (\ref{eqn51a}) we see that
\begin{subequations}
\label{limit1}
\begin{eqnarray}
&&u_+\approx\frac{1}{\sqrt{3}}\left(1+\frac{\sqrt{3}D}{2}\right),\\
&&v_+\approx\frac{1}{\sqrt{3}}\left(-1+\frac{\sqrt{3}D}{2}\right);
\end{eqnarray}
\end{subequations}
and
\begin{subequations}
\label{limit2}
\begin{eqnarray}
&&a\approx D/2,\,\,b\approx1,\\
&&\arctan\left(\frac{b}{a}\right)\approx\arctan\left(
\frac{2}{D}\right)\approx\frac{\pi}{2}.
\end{eqnarray}
\end{subequations}

\section{Comparison to analytic results from Ref.~\cite{imkeller}}

In this Appendix, we show that our results are consistent with those
obtained by Imkeller and Pavlyukevich from a different approach to
the barrier crossing problem of a particle exposed to L{\'e}vy stable
noise \cite{imkeller}.

We recall that in Ref.~\cite{imkeller}, the authors start from the formula
\begin{equation}
\langle\exp\left(ik\mathfrak{L}_{t}\right)\rangle=\exp\left(-\mathfrak{C}
(\alpha)t|k|^\alpha\right),
\end{equation}
with $0<\alpha<2$, where $\mathfrak{L}_t$ is the L{\'e}vy stable process
in the L{\'e}vy-Khintchin representation. The constant $\mathfrak{C}$ is
defined as
\begin{equation}
\label{CIP}
\mathfrak{C}(\alpha)=2\int_{0}^{\infty}{\frac{1-\cos y}{y^{1+\alpha}}{dy}}=
\frac{2\Gamma(1-\alpha)}{\alpha}\cos\left(\frac{\pi\alpha}{2}\right)
\end{equation}
for $\alpha$ strictly smaller than 2. In Ref.~\cite{imkeller}, the authors
investigate the system driven by the L{\'e}vy stable process $\varepsilon
\mathfrak{L}_t$, where $\varepsilon$ corresponds to the noise intensity:
\begin{equation}
\label{LProcIP}
\langle\exp\left(ik\varepsilon \mathfrak{L}_{t}\right)\rangle=\exp\left(
-\mathfrak{C}(\alpha)t\varepsilon^\alpha|k|^\alpha\right).
\end{equation}
Note that $\varepsilon$ is not completely equivalent to our noise intensity
$D$. Indeed, in the main body of the paper we investigate the system as driven
by the L{\'e}vy stable process $L_t$, with
\begin{equation}
\label{LProcOur}
\langle\exp\left(ikL_t\right)\rangle=\exp\left(-Dt|k|^\alpha\right).
\end{equation}
Since $L_t=\varepsilon\mathfrak{L}_t$, from comparison of Eqs.~(\ref{LProcOur})
and (\ref{LProcIP}), we see that
\begin{equation}
\label{DEpsConnection}
D=\mathfrak{C}(\alpha)\varepsilon^\alpha.
\end{equation}

Consider the exemplaric case of the truncated harmonic potential
\begin{equation}
U(x)=\left\{\begin{array}{ll}x^2/2, & -1\le x\le 1\\[0.2cm]
0 , & \mbox{otherwise}\end{array}\right..
\end{equation}
From our formula $T_{\mathrm{esc}}\simeq C(\alpha)/D^{\mu(\alpha)}$, assuming
that $\mu(\alpha)\approx 1$ (that we found to be nicely fulfilled for small and
intermediate values of $\alpha$, see Fig.~\ref{MU}), we find that
\begin{equation}
\label{OurT}
T_{\mathrm{esc}}\simeq C(\alpha)/D.
\end{equation}

The result reported in Ref.~\cite{imkeller} for this potential profile at
small noise intensities behaves like
\begin{equation}
\label{IPTesc}
\mathfrak{T}\approx\frac{\alpha}{2\varepsilon^\alpha},
\end{equation}
see Eq.~(24) from Ref.~\cite{imkeller}, with their constants $a$ and $b$
chosen as unity, $a=b=1$.

Thus, if we require $T_{\mathrm{esc}}=\mathfrak{T}$, then by virtue of
Eqs.~(\ref{OurT}), (\ref{IPTesc}), (\ref{DEpsConnection}), and (\ref{CIP}),
we find the relation
\begin{equation}
C(\alpha)=\frac{\alpha}{2}\mathfrak{C}(\alpha)=\Gamma(1-\alpha)\cos
\left(\frac{\pi\alpha}{2}\right).
\end{equation}
From this relation we find the asymptotic behavior of $C(\alpha)$,
\begin{equation}
C(\alpha)\approx\left\{\begin{array}{ll}
\Gamma(1)=1, & \alpha\rightarrow 0,\\[0.2cm]
\pi/2, & \alpha=1\\[0.2cm]
1/(2-\alpha)\rightarrow\infty, & \alpha\rightarrow2^-
\end{array}\right.,
\end{equation}
that agrees nicely with our numerical results, see Fig.~\ref{MU} in the
inset of the bottom graph corresponding to the truncated harmonic potential.

\begin{figure}
\includegraphics[width=8.6cm]{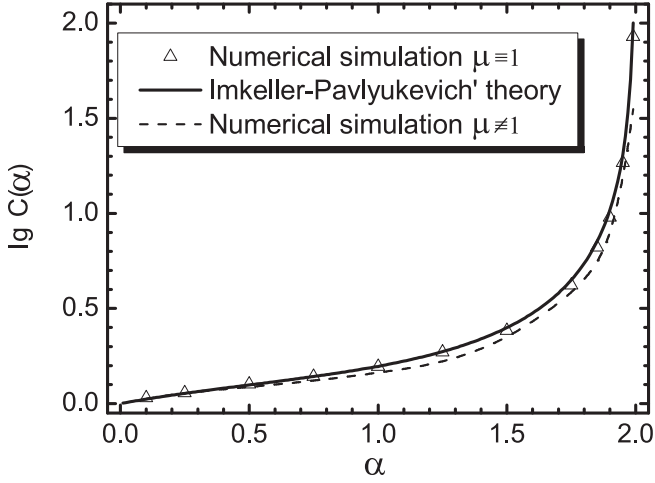}
\caption{\label{FigCompImk}Comparison of the coefficients $C(\alpha)$
obtained by using numerical simulations with the analytical behaviour
of $\mathfrak{C}(\alpha)$ (full line) derived in Ref.~\cite{imkeller},
for the truncated harmonic potential.
While the triangles denote our results for $\mu=1$ fixed, the dashed
line refers to our results with $\mu(\alpha)$, where we permit an
explicit dependence on $\alpha$.}
\end{figure}

Fig.\ref{FigCompImk} shows a direct comparison between the analytical
result from Ref.~\cite{imkeller} and our numerical results, by enforcing
the $\mu(\alpha)=1$ equality. The agreement is excellent.
We also show that taking a variable $\mu(\alpha)$ into account, only a slight
deviation is obtained.

\begin{figure}
\includegraphics[width=8.6cm]{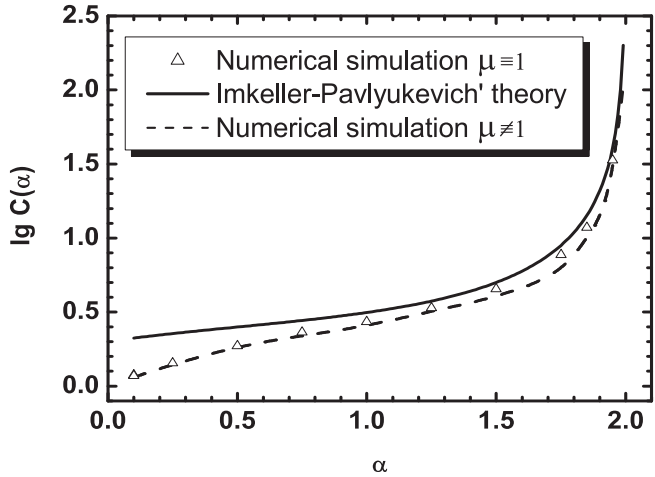}
\caption{\label{f5} Comparison of the coefficients $C(\alpha)$
obtained by using numerical simulations with the analytical behaviour
of $\mathfrak{C}(\alpha)$ (full line) derived in Ref.~\cite{imkeller},
for the bistable potential.}
\end{figure}

For comparison, Fig.~\ref{f5} shows the analogous results for the
case of the bistable potential. While general agreement is quite good,
for L{\'e}vy index decreasing below $\alpha=1$, an increasing deviation is
observed.
We note that
according to the results from Ref.~\cite{imkeller}, the coefficient $C$ is
$\pi$ at $\alpha=1$, in accordance with our result (\ref{escape}).

\end{appendix}


\begin{thebibliography}{99}

\bibitem{kramers} H. A. Kramers, Physica A \textbf{7}, 284 (1940).

\bibitem{pontryagin} L. S. Pontryagin, A. A. Andronov, and A. A. Vitt,
Zh. Eksp. Teor. Fiz. (Soviet Phys. JETP) \textbf{3}, 165 (1933).

\bibitem{stratonovich} R. L. Stratonovich, Topics of the Theory of Random
Noise, Vol.II (Gordon and Breach, New York, 1967).

\bibitem{haenggi} P. H{\"a}nggi, P. Talkner, and M. Bokrovec, Rev. Mod.
Phys. \textbf{62}, 251 (1990).

\bibitem{melnikov} V. I. Mel'nikov, Phys. Rep. \textbf{209}, 1 (1991).
Reaction rate problems

\bibitem{chandrasekhar} S. Chandrasekhar, Rev. Mod. Phys. \textbf{15},
1 (1943).

\bibitem{risken} H. Risken, The Fokker-Planck equation (Springer-Verlag,
Berlin, 1989).

\bibitem{bouchaud} J.-P. Bouchaud and A. Georges, Phys. Rep. \textbf{195},
127 (1990).

\bibitem{report1} R. Metzler and J. Klafter, J. Phys. A \textbf{37}, R161
(2004).

\bibitem{shlekla} J. Klafter and M. F. Shlesinger, Proc. Natl. Acad. Sci.
U.S.A. \textbf{83}, 848 (1986); A. Blumen, J. Klafter, and G. Zumofen, in
Optical Spectroscopy of Glasses, edited by I. Zschokke (Reidel, Amsterdam,
1986).

\bibitem{vlad} M. O. Vlad, R. Metzler, T. F. Nonnenmacher, and M. C. Mackey,
J. Math. Phys. \textbf{37}, 2279 (1996).

\bibitem{grote} R. F. Grote and J. T. Hynes, J. Chem. Phys. \textbf{73},
2715 (1980).

\bibitem{haenggi1} P. H{\"a}nggi and F. Mojtabai, Phys. Rev. A \textbf{26},
1168 (1982).

\bibitem{cpl} R. Metzler and J. Klafter, Chem. Phys. Lett. \textbf{321}, 238
(2000).

\bibitem{report} R. Metzler and J. Klafter, Phys. Rep. \textbf{339}, 1
(2000).

\bibitem{scher} H. Scher and E. W. Montroll, Phys. Rev. B \textbf{12}, 2455
(1975); G. Pfister and H. Scher, Phys. Rev. B \textbf{15}, 2062 (1977).

\bibitem{klablushle} J. Klafter, A. Blumen, and M. F. Shlesinger, Phys. Rev.
A \textbf{35}, 3081 (1987).

\bibitem{hughes} B. D. Hughes, Random Walks and Random Environments, Vol. 1:
Random Walks (Oxford University Press, Oxford, 1995).

\bibitem{plasma}
V. Y. Gonchar, A. V. Chechkin, E. L. Sorokovoi, V. V.
Chechkin, L. I. Grigor'eva, and E. D. Volkov, Plasma Phys. Rep. {\bf 29},
380 (2003); O. G. Bakunin, Plasma Phys. Rep. {\bf 29}, 955 (2003);
R. Jha, P. K. Kaw, D. R. Kulkarni, and J. C. Parikh, Phys. Plasmas \textbf{10},
699 (2003); Y. Marandet, H. Capes, L. Godbert-Mouret, M. Koubiti,
J. Rosato, and R. Stamm, Euro. Phys. J. D \textbf{39}, 247 (2006).

\bibitem{looping} I.~M. Sokolov, J. Mai and A. Blumen,
Phys. Rev. Lett. {\bf 79}, 857 (1997);
D.~Brockmann and T.~Geisel, Phys.~Rev.~Lett.~\textbf{91}, 048303;
M. A. Lomholt, T. Ambj{\"o}rnsson, and R. Metzler, Phys. Rev. Lett.
\textbf{95}, 260603 (2005).

\bibitem{walter} H. Katori, S. Schlipf, and H. Walther,
Phys. Rev. Lett. {\bf 79}, 2221 (1997).

\bibitem{sms} G. Zumofen and J. Klafter, Chem. Phys. Lett. {\bf 219} 303
(1994); E. Barkai and R. Silbey, Chem. Phys. Lett. {\bf 310}, 287 (1999).

\bibitem{ditlevsen} P. D. Ditlevsen, Geophys. Res. Lett. \textbf{26}, 1441
(1999).

\bibitem{levy} P. L{\'e}vy, Th{\'e}orie de
l'addition des variables al{\'e}atoires (Gauthier-Villars, Paris, 1954).

\bibitem{gnedenko} B. V. Gnedenko and A. N. Kolmogorov,
Limit Distributions for Sums of Random Variables
(Reading, MA: Addison-Wesley, 1954).

\bibitem{imkeller} P. Imkeller and I. Pavlyukevich, J. Phys. A \textbf{39},
L237 (2006); P. Imkeller and I. Pavlyukevich, Stochast. Proc. Applic.
\textbf{116}, 611 (2006).

\bibitem{ditlevsen1} P. D. Ditlevsen, Phys. Rev. E \textbf{60}, 172 (1999).

\bibitem{epl} A. V. Chechkin, V. Yu. Gonchar, J. Klafter, R. Metzler,
Europhys. Lett. \textbf{72}, 348 (2005).

\bibitem{bao} J.-D. Bao, H.-Y. Wang, Y. Jia, Phys. Rev. E \textbf{72},
051105 (2005).

\bibitem{plasma1} A. V. Chechkin, V. Y. Gonchar, and M. Szydlowsky, Phys.
Plasma {\bf 9}, 78 (2002); A. V. Chechkin and V. Yu. Gonchar, Zh. Eksp.
Teor. Fiz. (JETP) \textbf{118}, 3, 730 (2000).

\bibitem{REMM} The simulation was performed in two ways: first, we used the
programming language Borland C++ Builder 6 and, second, the symbolic
mathematics package Mathematica 5. The step of time was taken equal to 0.01
in Borland C++ Builder program and 0.1 in simulation with Mathematica. The
calculations made on Borland C++ Builder and Mathematica gave the same results
(the accuracy was better than 0.7 percent). It was also proved that the
simulation scheme is independent of the time quantization parameter (the
inaccuracy did not exceed an error for the usual scheme of integrating an
ordinary differential equation of first order by using the method of
rectangles).

\bibitem{REMM1} We note that in Fig. \ref{MET} the values of the noise intensity
$D$ for the Gaussian case is larger than the barrier height and thus
Eq.~(\ref{eq4}) is not valid. Therefore, we here compare the result of
numerical simulation with Eq.~(44) from the treatise by Malakhov
\cite{malakhov}, which allows us to avoid the assumption of the high barrier.
The corresponding theoretical curves are shown in Fig.~\ref{MET} as bold lines
for Levy index $\alpha=2$. Obviously, there is excellent agreement between
analytical solution given in Ref.~\cite{malakhov} and the simulation results.

\bibitem{malakhov} A. N. Malakhov, Chaos \textbf{7}, 488 (1997).

\bibitem{REMM2} For the values of $\alpha$ and $D$ shown in Fig.~\ref{FigPDF}
the simulations were performed twice and the results compared. A good agreement
was obained, proving the reliability of the simulation.

\bibitem{klimontovich} Yu. L. Klimontovich, Statistical Theory of Open
Systems, Vol. I (Kluwer Academic Publishers, Dordrecht, 1995).

\bibitem{REMM3} Note that the Fourier transform of Eq.~(\ref{conflu}) reads
\begin{equation}
j(k)=\int_{-\infty}^{\infty}j_0e^{ikx}dx=2\pi j_0\delta(k).
\end{equation}

\bibitem{chambers} J. M. Chambers et al., J. Americ. Statist. Assoc.
\textbf{71}, 340 (1976).

\bibitem{zolotarev} V. M. Zolotarev, Moscow Mir, p. 187 (1983).


\end{thebibliography}
\end{document}